\documentclass[aps,prb,twocolumn,showpacs,preprintnumbers]{revtex4-1}
\usepackage{graphicx}
\usepackage{bm}
\renewcommand{\v}[1]{{\bf #1}}

\def\be{\begin{eqnarray}}
\def\ee{\end{eqnarray}}
\newcommand{\nn}{\nonumber\\}

\renewcommand{\t}[1]{{\tilde #1}}

\newcommand{\gr}{{\nabla}}

\begin{document}

\title{Featureless Mott Insulators}
\author{Chyh-Hong Chern}
\email{chchern@ntu.edu.tw} \affiliation{Department of Physics, Center for Theoretical Sciences, and Center for Quantum Science and Engineering, National Taiwan University, Taipei 10617, Taiwan}

\begin{abstract}
A family of the pair hopping models exhibiting the incompressible quantum liquid at fractional filling $1/m^D$ is constructed in $D$ dimensional lattice.  Except in one dimension, the lattice is the generalized edge-shared triangular lattice, for example the triangular lattice in two dimensions and tetrahedral lattice in three dimensions.  They obey the new symmetry, conservation of the center-of-mass position proposed by Seidel et al..\cite{Seidel2005}  The uniqueness of the ground state is proved rigorously in the open boundary condition.  The finiteness of the excitation energy is calculated by the single mode approximation.  \end{abstract}

\maketitle

\section{Introduction}\label{section:introduction}

The featureless Mott insulator blocks the charge transportation due to the strong electron-electron interaction and the ground state exhibits no symmetry breaking.  Their existence is very rare both in the experimental systems and in the theoretical models.  There are two examples in the two dimensional systems.  One is the famous fractional quantum Hall effect, where the ground state is an incompressible liquid when the filling factor is $1/q$ ($q$ = odd integer).  The other example is the quantum dimer model in the two dimensional triangular lattice, where the ground state is a disordered dimer liquid that separates the excitations by a finite energy gap.  

Very interestingly, Seidel et al. pointed out that these two systems actually belong to the same type of Hamiltonian.\cite{Lee2004PRL,Seidel2005}  Namely, their Hamiltonians preserve both the center-of-mass momentum and the center-of-mass position.  They showed that the fractional quantum Hall system in the lowest Landau level on the torus described by the following pseudo-potential Hamiltonian\cite{trugman1985PRB}
\be 
H=\int d^2r
d^2r' \gr^2\delta(\v r-\v r') \psi^+(\v r)\psi^+(\v r')\psi(\v
r')\psi(\v r)\label{pseudo} 
\ee 
can map to a pair hopping model in one dimensional lattice
 \be
&&H=\sum_{R,x,y} f^*(x)f(y) C^\dagger_{R+x}C^\dagger_{R-
x}~C_{R-y}C_{R+y}\nn&&f(x)=\kappa^{3/2}\sum_{n} (x-nL)
e^{-\kappa^2(x-nL)^2} \label{pairhopp}, 
\ee 
where  \be L=L_xL_y/2\pi
l_B^2,~~~ \kappa=2\pi l_B/L_y,\label{redef} \ee $L_x$ and
$L_y$ are the linear dimensions of the torus, and $l_B=\sqrt{\hbar
c/eB}$ is the magnetic length.   It can be obviously seen that Eq.(\ref{pairhopp}) describes the hopping that preserves the center-of-mass position.  Two electrons annihilated at $R\pm y$ hop to $R\pm x$ with the effective hopping range $1/\kappa$ and center-of-mass position preserves at $R$.  Due to this beautiful property, the ground states of the $1/q$ fractional quantum Hall liquid with the $q$-fold degeneracy can be labeled by the $q$ different center-of-mass positions.  Furthermore, Seidel et al. also showed that the ground state of Eq.(\ref{pairhopp}) is a charge density wave with the amplitude $\sim
e^{-c/\kappa^2}\label{cdw}$ where $c$ is a constant of O(1), and the energy gap is finite for any finite $\kappa$.  For very small $\kappa$, where Eq.(\ref{pairhopp}) becomes the long range hopping model, the charge density wave amplitude is exponentially small.  One can safely say that the ground state describes a \emph{featureless Mott insulator} without local order parameter.

The new symmetry of the center-of-mass position conservation paves a new way to search the featureless Mott insulator.  Recently, inspired by the higher dimensional generalization of the quantum Hall effect\cite{Zhang2001Science} and the Haldane's pseudopotential Hamiltonian,\cite{Haldane1983PRL} Chern et al. constructed a model for the incompressible liquid in the two-dimensional triangular lattice.\cite{Chern2007PRL}  They showed that the ground state is unique without local order parameter.  They also computed the excitation gap using the single mode approximation.  This paper will serve as the extended version of that Letter.  For this purpose, we organize the paper in the following:  In the section \ref{section:su2}, we review the Haldane construction of the pseudopotential method on the two sphere.  We will show explicitly that the Haldane pseudopotential Hamiltonian can map to a long-range pair hopping model with the conservation of the center of mass position.  In section \ref{section:su3}, we provide the detail calculation of our previous Letter and generalize it to the SU(N) model.  In section \ref{section:sma}, we provide the detail calculation of the single mode approximation in the SU(3) case.  Finally, we conclude and summarize in the section \ref{section:dis_con}.  We also include several appendixes for the readers to follow the group-theoretical method easily.

\section{One-dimensional lattice model for the incompressible quantum liquid} \label{section:su2}

The quantum Hall effect can be considered on the two sphere subject to the uniform
magnetic field by the U(1) magnetic monopole at the
center.\cite{Haldane1983PRL}  In this case, the Laughlin wave function becomes the exact ground state of the Eq.(\ref{pseudo}).  We will show that it can map to a pair hopping model with the conservation of the center-of-mass position and long range hopping integral.

In the presence of the U(1) magnetic monopole flux, the
single-particle wavefunction is described by the monopole vector
spherical harmonics\cite{Tamm1931, Wu1976NPB} which can be denoted
by the SU(2) $|l,m>$ state, where $l$ can be integers or
half-integers and $m$ is the magnetic quantum number. The Landau
level spectrum is given by\cite{Haldane1983PRL}
\begin{eqnarray}
E_k=\frac{\hbar^2}{2MR^2}l(l+1) \label{Eq:ll_energy}
\end{eqnarray}
where $M$ is the mass of the electrons and $R$ is the radius of
the sphere.  If the total magnetic flux is $2S$, $l=S+k$, where
$k$ is the Landau level index.  Because $k$ is only an integer,
$S$ is either integer or half-integer.  Each Landau level has
$2(S+k)+1$ degeneracy. In the lowest Landau level, $k=0$, the single-particle
wavefunction can be written as
\begin{eqnarray}
\psi_i = \sqrt{\frac{(2S)!}{(S+i)!(S-i)!}}u^{S+i}v^{S-i}
\label{Eq:su2_coherent_state}
\end{eqnarray}
where $i=-S,-S+1,..,S-1,S$ and the $(u,v)$ is the two-dimensional
complex spinor given by
\begin{eqnarray}
\left(\begin{array}{c} u \\ v \end{array}\right)=\left(
\begin{array}{l} \cos\frac{\theta}{2}e^{-i\phi/2} \\
\sin\frac{\theta}{2}e^{i\phi/2}\end{array}\right)
\label{Eq:su2_spinor}
\end{eqnarray}
where $\theta$ and $\phi$ parameterizing the sphere are known as
the polar and azimuthal coordinates.  This is the special property
of the lll that the single-particle wavefunction can be completely
described only by one quantum number $i$. Therefore, one can treat
the configuration space as the 1D chain with the number of site
$2S+1$ and the lattice site is labelled by $i$.  If the filling
factor $\nu =1$, the number of the particles $N=2S+1$. Then, the
many-body wavefunction is the Slater determinant given by
\begin{eqnarray}
\Psi = \prod_{k<l}^N(u_kv_l-u_lv_k) \label{Eq:su2_IQHE_state}
\end{eqnarray}
where $k$ and $l$ are the particle indices. It is easy to see that
Eq.(\ref{Eq:su2_IQHE_state}) is the unique many-body fermionic
wavefunction for $\nu=1$.

The quantum Hall state with filling factor $\nu=1/m$ celebrated as
the Laughlin wavefunction\cite{Laughlin1983PRL} can be written
as\cite{Haldane1983PRL}
\begin{eqnarray}
\Psi^m=\prod_{k<l}^N(u_kv_l-u_lv_k)^m \label{Eq:su2_FQHE_state}
\end{eqnarray}
where $m$ is an odd integer.  In Eq.(\ref{Eq:su2_FQHE_state}), the maximum power of $u_k$
becomes $2mS$, indicating that the single-particle wavefunction given by the Eq.(\ref{Eq:su2_coherent_state}) is in the spin-$mS$ state and therefore the dimension of the single- particle states (or say the lattice) is $2mS+1$.   While keeping the number of
particles $N=2S+1$ the same, Eq.(\ref{Eq:su2_FQHE_state})
describes a state with filling factor
\begin{eqnarray}
\nu=\frac{2S+1}{2mS+1}, \label{Eq:f_filling}
\end{eqnarray}
which goes to $1/m$ as $S$ goes to infinity.  Because $S$ scales
as $R^2$ shown in the Eq.(\ref{Eq:ll_energy}), infinite $S$
indicates to take the thermodynamic limit.

Eq.(\ref{Eq:su2_FQHE_state}) is the unique ground state of the
following many-body Hamiltonian
\begin{eqnarray}
H_1=\frac{1}{2}\sum_{(ij)}\sum_{k=1,\ \text{odd}}^{k\leq m-2}
\alpha_k P^{2mS-k}_{ij} \label{Eq:su2_Hamiltonian_haldane}
\end{eqnarray}
where $\alpha_k$ are positive-definite and $P^{2mS-k}_{ij}$ are
the projection operators that project two spin-$mS$ states to the two-body states of total spin $2mS-k$ for the pair $(ij)$.  We note that $k$ can be only the odd integers
because the two-body $2mS-k$ states with odd $k$ are antisymmetric
upon particle exchanges. In Eq.(\ref{Eq:su2_FQHE_state}), the term
with the maximum power of $u_iu_j$ for any pair $(ij)$ is
\begin{eqnarray}
(u_iu_j)^{m(2S-1)}(u_iv_j-u_jv_i)^m,
\end{eqnarray}
which indicates that no two-body $2mS-k$ states for $k\leq m-2$
for any pair $(ij)$. Therefore, Eq.(\ref{Eq:su2_FQHE_state}) is
the zero energy state of Eq.(\ref{Eq:su2_Hamiltonian_haldane}). On
the other hand, because there is no two-body $2mS-k$ state for
$k\leq m-2$ for any pair $(ij)$, any ground state wavefunction
$\chi$ must have the following form
\begin{eqnarray}
\chi\sim f(u_1,v_1;...;u_N,v_N)\prod_{k<l}^N(u_kv_l-u_lv_k)^m,
\label{Eq:chi}
\end{eqnarray}
where $f$ is the symmetric function for any pair exchange.  For
the single-particle wavefunction to be described by $l=mS$, the
power of $u_k$ for each particle in Eq.(\ref{Eq:chi}) has to be
$2mS$.  However, the factor
\begin{eqnarray}
\prod_{k<l}^N(u_kv_l-u_lv_k)^m
\end{eqnarray}
in Eq.(\ref{Eq:chi}) already exhausts the quota of the power of
$u_k$. Then, $f$  can only be a constant.  $\chi\sim \Psi_m$ and
Eq.(\ref{Eq:su2_FQHE_state}) is indeed the unique ground state of
Eq.(\ref{Eq:su2_Hamiltonian_haldane}).

The SU(2) spin model of Eq.(\ref{Eq:su2_Hamiltonian_haldane}) can
be formulated as the lattice hopping model.  As we mentioned, the
single-particle state can be labelled by one quantum number $i$.
Let us denote
\begin{eqnarray}
|mS, i \rangle \to c^\dag_i |0\rangle \label{Eq:su2_second_quan}
\end{eqnarray}
Then, Eq.(\ref{Eq:su2_Hamiltonian_haldane}) can be written as
\begin{eqnarray}
\mathcal{H}_1=\alpha\sum_{n}\sum_{k,q} F(n,k,q) c^\dag_k
c^\dag_{n-k}c_{n-q}c_q. \label{Eq:su2_H_hopping}
\end{eqnarray}
for $m=3$, where $F(n,k,q)$ can be obtained by the SU(2)
Clebsh-Gordan coefficient
\begin{eqnarray}
&&F(n,k,q) = f^*(n,q)f(n,k), \nonumber\\
&&f(n,k) =
\frac{(-1)^{2n}\sqrt{3S}(2k-n)\sqrt{12S-1}}{\sqrt{(3S-k)(3S+k)(3S+k-n)}}
\nonumber \\ &\times&
\frac{\sqrt{(6S-n-1)!(6S+n-1)!}}{\sqrt{(3S-k+n)(12S-1)!(3S-k-1)!}}
\nonumber \\&\times&
\frac{\sqrt{(6S-1)!}}{\sqrt{(\!3S\!+\!k\!-\!1\!)!(\!3S\!+\!k\!-\!n\!-\!1\!)!(\!3S\!-\!k\!+\!n\!-\!1\!)!}}
\label{Eq:su2boson}
\end{eqnarray}
It can be easily seen that Eq.(\ref{Eq:su2_H_hopping}) is a
center-of-mass conserving hopping. Two electrons annihilated with
the center-of-mass position $\frac{1}{2}(k+n-k)=\frac{n}{2}$ will
be created in pair with the same center-of mass position.  The
reason that the hopping matrix element $F$ is $n$-dependent is due
to the presence of the open boundary.  Although $F$ has long range hopping, the hopping matrix elements is exponentially small as the relative
distance is comparable to the system size.

\begin{figure}[htb]
    \includegraphics[width=0.45\textwidth]{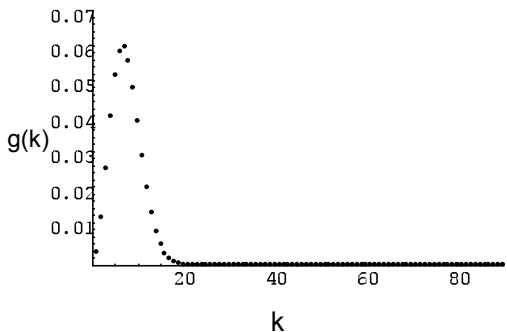}
\caption{The $g(k)$ versus $k$ for $S=30$.}
\label{Fig:hoping_su2_ss}
\end{figure}

\begin{figure}[htb]
    \includegraphics[width=0.45\textwidth]{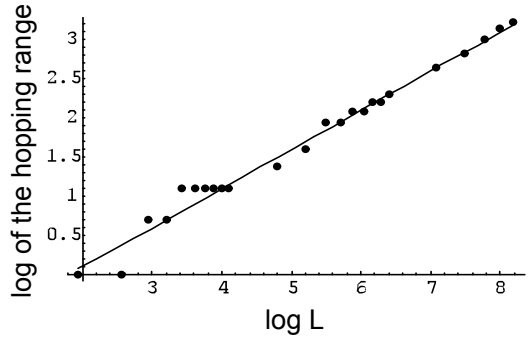}
\caption{A Log-Log plot of the width in $|g(k)|^2$ as a function
of $L=6S+1$. We have calculated for
S=0,1,2,..,10,20,30,..,100,200,300,..,600.}
\label{Fig:hoping_su2_width}
\end{figure}

To analysis hopping range, let us consider the hopping with the
center-of-mass to be at the origin, namely $n=0$.  Then
$f(n,k)=g(k)$, where
\begin{eqnarray}
g(k)=\frac{2k\sqrt{3S}((6S-1)!)^2}{(k^2\!-\!9S^2)\sqrt{(12S\!-\!2)!}(3S\!-\!k\!-\!1)!(3S\!+\!k\!-\!1)!}.
\end{eqnarray}
$g(k=0)=0$ indicates that no two electrons can be created
(annihilated) at the same sites. For $S=30$, $g(k)$ is calculated
in the Fig.(\ref{Fig:hoping_su2_ss}).  Taking $k=3S$,
\begin{eqnarray}
g(3S)\sim \sqrt{S}e^{-6S\log 2},
\end{eqnarray}
which has an exponentially small tail.
A proper definition of the hopping range can be defined as the half
width of the hump of the $|g(k)|^2$.  The result that the hopping
range scales with $S$ is given in the
Fig.(\ref{Fig:hoping_su2_width}), where we show the log-log
relation between the hopping range and the size of the system
$L=6S+1$.  The trend line in the Fig.(\ref{Fig:hoping_su2_width})
is the best fit given by $\frac{1}{2}\log L-0.9$, which indicates that
the hopping range scales as $L^{1/2}$.

In summary, using the SU(2) weight space, we have considered the 2D fractional
quantum Hall effect as a 1D lattice model with the open boundary
condition.  The weight number, labeling the lattice site, is the
$z$-component coordinate of the coherent state on the sphere. That
the fractional quantum Hall state does not exhibit any
density-wave order on the sphere implies that the ground state has
no long range order in the 1D lattice model. On the other hand,
the sphere has no edge. The energy spectrum of the
Eq.(\ref{Eq:su2_H_hopping}) does not include the edge modes that
occur in the open 2D plane system. Thus, there is no gapless
excitation corresponding in this model.  Note that although the fractional quantum Hall state is the
incompressible liquid shown in the experiments, theoretically the finiteness of the excitation gap is only proved by the single mode approximation\cite{girvin1985, girvin1986}.

\section{Higher dimensional lattice model for the incompressible quantum liquid} \label{section:su3}
There is a natural generalization of
Eq.(\ref{Eq:su2_coherent_state}) to the SU(N) coherent state,
which is labelled by $(N-1)$ quantum numbers \emph{without
degeneracy}.  The SU(N) coherent state is known as the SU(N)
$(p,0,..,0)$ multiplet.  Similar to the SU(2) case, we will
construct a family of the faithful lattice models for the
incompressible quantum liquid using the SU(N) coherent states.  We shall start with the SU(3) for the pedagogical purpose.

\begin{figure}[htb]
\centerline{\includegraphics[angle=0,scale=0.3]{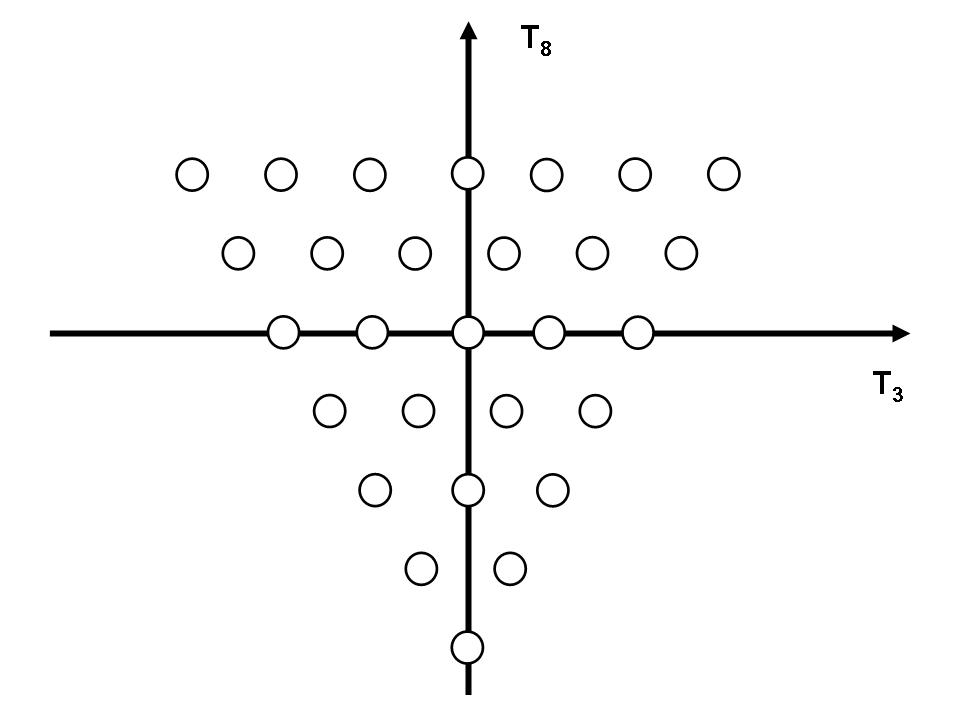}}
\caption{The weight space of $(6,0)$. $T_3$ and $T_8$ are the
Gel-Mann matrices forming the Cartan subalgebra.}
\label{Fig:weight_su3_6_0}
\end{figure}

The SU(3) coherent states are given by
\begin{eqnarray}
\sqrt{\frac{p!}{m_1!m_2!m_3!}}u^{m_1}v^{m_2}w^{m_3}, \
m_1\!+\!m_2\!+\!m_3\!=\!p \label{Eq:su3_coherent_state}
\end{eqnarray}
which forms the multiplet described the SU(3) $(p,0)$
representation and $m_i$ are integers. The $(u,v,w)$ in
Eq.(\ref{Eq:su3_coherent_state}) is the complex spinor which can
be represented by
\begin{eqnarray}
\psi_\alpha(z_1,z_2)=\left( \begin{array}{c} u \\ v \\ w \end{array}\right)=\frac{1}{\sqrt{1+\bar{z}_1z_1+\bar{z}_2z_2}}\left(\begin{array}{c}1 \\
z_1
\\ z_2 \end{array} \right), \label{Eq:su3_uvw}
\end{eqnarray}
where $z_i$ are the complex numbers to parameterize CP$^2$ and $\alpha$ is the spinor index.  As shown in the Appendix, the SU(3) $(p,0)$ states is the lowest
Landau level in the quantum Hall problem in CP$^2$.  The SU(3)
Cartan subalgebra contains two generators $T_3$ and $T_8$.
 The states of Eq.(\ref{Eq:su3_coherent_state}) acquire the coordinates in the $T_3$ and $T_8$ space, which form the two-dimensional triangular lattice shown in the
Fig.\ref{Fig:weight_su3_6_0}.
The number of sites of the two-dimensional lattice representing
the SU(3) $(p,0)$ multiplet is given by
\begin{eqnarray}
d(p)=\frac{1}{2}(p+1)(p+2) \label{Eq:su3_dimension}
\end{eqnarray}
If the number of electrons $N=d(p)$, the many-body fermionic
wavefunction is the Slater determinant given by
\begin{eqnarray}
\Psi = \left| \begin{array}{ccccc} u_1^p & u_1^{p-1}v_1 & . & . &
w_1^p \\ u_2^p & u_2^{p-1}v_2 & . & . & w_2^p \\ . & . & . & . & .
\\ . & . & . & . & . \\ u_N^p & u_N^{p-1}v_N & . & . & w_N^p
\end{array} \right| \label{Eq:su3_IQHE}
\end{eqnarray}
up to the normalization constant.  It is also not hard to see that
Eq.(\ref{Eq:su3_IQHE}) is the unique fermionic many-body
wavefunction when $N=d(p)$.  Next, let us consider the natural
generalization of Eq.(\ref{Eq:su2_FQHE_state}) written by
\begin{eqnarray}
\Psi^m = \left| \begin{array}{ccccc} u_1^p & u_1^{p-1}v_1 & . & .
& w_1^p \\ u_2^p & u_2^{p-1}v_2 & . & . & w_2^p \\ . & . & . & . &
.  \\ . & . & . & . & . \\ u_N^p & u_N^{p-1}v_N & . & . & w_N^p
\end{array} \right|^m \label{Eq:su3_FQHE}
\end{eqnarray}
Because the highest power of $u$ for each particle in
Eq.(\ref{Eq:su3_FQHE}) is $mp$, the new coherent state is
described by the SU(3) $(mp,0)$ multiplet.  Therefore,
Eq.(\ref{Eq:su3_FQHE}) is the many-body state with the filling
factor
\begin{eqnarray}
\nu=\frac{(p+1)(p+2)}{(mp+1)(mp+2)} \label{Eq:su3_filling_factor}
\end{eqnarray}
which becomes $1/m^2$ in the thermodynamic limit.  We propose the
following SU(3) spin Hamiltonian
\begin{eqnarray}
H=\frac{1}{2}\sum_{i\ne j}\sum_{q=1, \ \text{odd}}^{q\leq m\!-\!2}
\kappa_q~P_{ij}^{(2mp\!-\!2q,q)}. \label{Eq:su3_H}
\end{eqnarray}
where the operator $P_{ij}^{(2mp\!-\!2q,q)}$ operates on the
direct product states of two spins $i$ and $j$ and projects them
onto the $(2mp-2q,q)$ states, and $\kappa_q$ are
positive-definite.  The SU(3) two-spin states are the
generalization of the SU(2) angular momentum addition.  The direct
product of two SU(3) multiplets can be also block-diagonalized
such that each block is described a SU(3) multiplet denoted by two
integers $(a,b)$.  In this case, the direct product of two SU(3)
$(mp,0)$ multiplets can computed as
\begin{eqnarray}
(mp,0)\times (mp,0)|_{\text{a}} = \bigoplus_{q=1,
\text{odd}}^{mp}(2mp-2q,q) \label{Eq:su3_CG}
\end{eqnarray}
where the subscript "a" denotes the antisymmetric combination and
$k$ are the odd integers.  Furthermore, every SU(3) multiplet can
be also block-diagonalized by its SU(2) subgroup.  In our case,
the SU(3) $(p,0)$ multiplet can be decomposed as
\begin{eqnarray}
(p,0)=\bigoplus_{k=0}^p \frac{k}{2}\label{Eq:su3_su_2decom}
\end{eqnarray}
which can be easily checked by the counting the dimensionality.

The Eq.(\ref{Eq:su3_FQHE}) is the zero-energy state of
Eq.(\ref{Eq:su3_H}).  We first look at the two-spin state of
Eq.(\ref{Eq:su3_IQHE}).  Because all sites are occupied, the SU(3)
two-spin state with the maximum SU(2) quantum number is the
$(2p-2,1)$ multiplet.  Because Eq.(\ref{Eq:su3_FQHE}) is the
($m$-fold) product of the Eq.(\ref{Eq:su3_IQHE}), the SU(3) multiplet
with the maximum SU(2) quantum number that the direct product of
$m$ $(2p-2,1)$ multiplets can yield is $(2mp-2m,m)$.  Therefore,
due to the SU(3) symmetry, the two-spin states in
Eq.(\ref{Eq:su3_FQHE}) does not contain the multiplets
$(2mp-2k,k)$ for $k\leq m-2$.  The Hamiltonian is the SU(3)
generalization of the Eq.(\ref{Eq:su2_Hamiltonian_haldane}).  Our
generalized Laughlin wavefunction of Eq.(\ref{Eq:su3_FQHE}) is the
zero-energy state of the generalized Hamiltonian of
Eq.(\ref{Eq:su3_H}).  Moreover, it is also the \emph{unique}
ground state. In the Letter\cite{Chern2007PRL}, we have
demonstrated rigorously that Eq.(\ref{Eq:su3_FQHE}) is the
\emph{unique} ground state of the Hamiltonian of
Eq.(\ref{Eq:su3_H}).  Some supplemental details will be discussed
in the Appendix.

Eq.(\ref{Eq:su3_H}) can be written as a lattice hopping model.
Similar to Eq.(\ref{Eq:su2_second_quan}), the electron creation
operator can be defined as
\begin{eqnarray}
|(mp,0);j,j_3>\to c^\dag_{(j,j_3)}|0>, \label{Eq:su3_second_quan}
\end{eqnarray}
where the states are denoted by the quantum numbers of the SU(2)
subgroup. This coordinate system is equivalent to the quantum
numbers of $T_3$ and $T_8$, since there is no degeneracy on the
lattice.  Using Eq.(\ref{Eq:su3_second_quan}), the Hamiltonian for
$m=3$ can be expressed by
\begin{eqnarray}
H&&=\kappa
\sum_{j,L,L_3}\sum_{l,l_3}\sum_{k,k_3}F^{j,L,L_3}_{l,l_3}F^{j,L,L_3}_{k,k_3}\nonumber\\&&
c^\dag_{(l,l_3)}c^\dag_{(j-l+\frac{1}{2},L_3-l_3)}
c_{(j-k+\frac{1}{2},L_3-k_3)}c_{(k,k_3)},\label{Eq:su3_php}
\end{eqnarray}
where $\kappa$ is a positive number and $F^{j,L,L_3}_{l,l_3}$ is
the SU(3) Clebsh-Gordan coefficient from two $(3p,0)$ multiplets
to the $(6p-2,1)$ subspace shown in Eq.(\ref{Eq:su3_CG}). In
Eq.(\ref{Eq:su3_php}), the center-of-mass position is conserved at
$(j+1/2,L_3)$ in the pair-annihilation and pair-creation process.

\begin{figure}[htb]
\centerline{
    \includegraphics[width=0.4\textwidth]{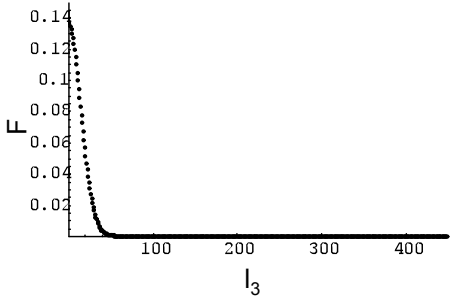}}
\caption{$F^{p,3p-\frac{3}{2},3p-1,0}_{\frac{3p}{2},l_3}$,
$p=300$, $l_3=1..\frac{3p}{2}-1$} \label{Fig:hoping_p300_su3}
\end{figure}

\begin{figure}[htb]
\centerline{
   \includegraphics[angle=0,scale=0.4]{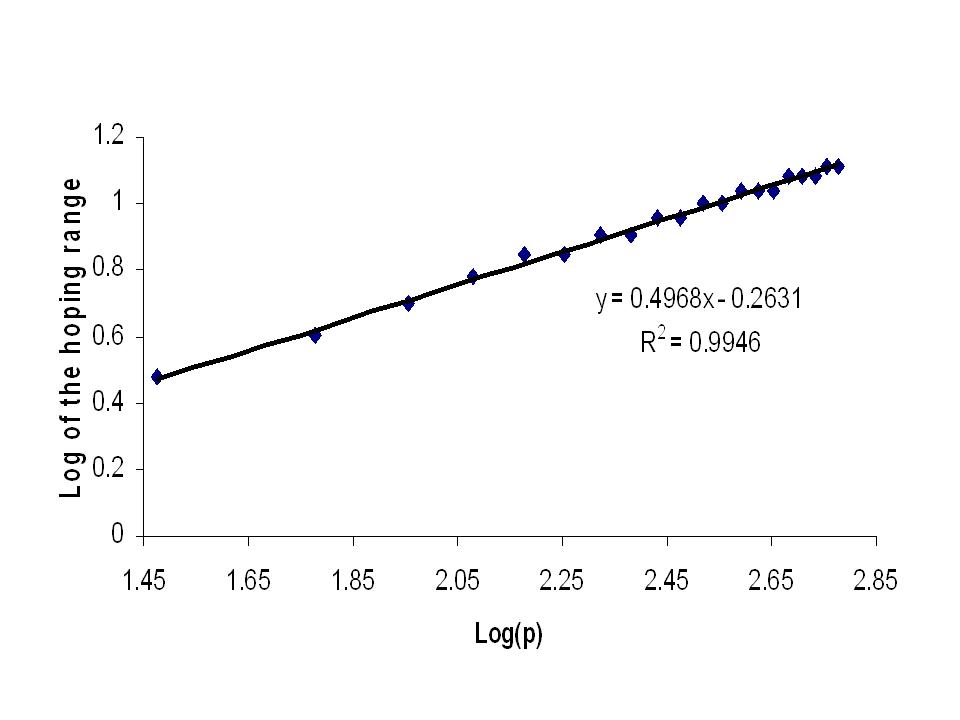}}
\caption{A log-log plot of the hopping range to $p$ from 30 to 600.
The straight line is the best fit.  The vertical axis is the log
of the hopping range and the horizontal one is $\log p$. The hopping
range scales as $p^{\frac{1}{2}}$ } \label{Fig:HopRangSu3}
\end{figure}

The Hamiltonian of Eq.(\ref{Eq:su3_php}) also describes a
long-ranged hopping process.  The $F^{j,L,L_3}_{l,l_3}$ can be
computed exactly as\cite{Rowe1997JMP,Rowe2000JMP}
\begin{eqnarray}
&&F^{j,L,L_3}_{l,l_3}=-\frac{(3p)!}{(6p)!}\sqrt{\frac{(6p\!-\!1)2!(6p\!+\!L\!-\!j\!-\!2l\!-\!\frac{1}{2})!}{(3p\!-\!2j\!+\!2l\!-\!1)!}}\nonumber\\\times&&\sqrt{\frac{(6p\!-\!L\!-\!j\!-\!\frac{3}{2})!(2j\!+\!1)!}{(3p\!-\!2l)!}}\frac{\sqrt{(2j-2l+2)!(2l+1)!}}{(2j-2l)!(2l-1)!}\nonumber
\\ \times&& (-1)^{\frac{1}{2}+j+L_3-2l}\sqrt{2L+1}\left(\begin{array}{ccc}j+\frac{1}{2}-l & l & L \\ L_3-l_3 & l_3 &
L_3\end{array}\right)\nonumber
\\
\times&\Big(&\frac{1}{2l}\left\{\begin{array}{ccc}\frac{1}{2} & 0
& \frac{1}{2}\\t & l & j \\t\!+\!\frac{1}{2} & l &
L\end{array}\right\}\!-\!\frac{1}{2t\!+\!1}\left\{\begin{array}{ccc}0
& \frac{1}{2} & \frac{1}{2} \\ t\!+\!\frac{1}{2} &
l\!-\!\frac{1}{2} & j \\ t\!+\!\frac{1}{2} & l &
L\end{array}\right\}\!\Big),\label{Eq:su3_F_function}
\end{eqnarray}
where $t=j-l$ and the Wigner 6-$j$ and 9-$j$ symbols are
explicitly used.  In Fig.\ref{Fig:hoping_p300_su3}, we plot the
$F^{j,L,L_3}_{l,l_3}$ for $p=300$, $j=3p-\frac{3}{2}$, $L=3p-1$,
and $l=\frac{3p}{2}$ as the function of $l_3$.
It is the pair-hopping integral with the relative distance
$\sqrt{3+4l_3^2}$.  Because two electrons are not on the same
rows, $F^{j,L,L_3}_{l,l_3}$ is not zero at $l_3=0$.  The hopping
range can be also defined as the half-width of $|F|^2$.  In
Fig.\ref{Fig:HopRangSu3}, we show the log-log relation between the
hopping range and $p$.  The result suggests that the hopping range
scales as $\sqrt{p}$.   Similar to the SU(2) case, when the
relative distance is the same order of the system size, the
pair-hopping integral is exponentially decayed as
\begin{eqnarray}
F^{p,3p-\frac{3}{2},3p-1,0}_{\frac{3p}{2},\frac{3p}{2}} \sim \
\sqrt{6p(6p-1)}\ e^{-3p\log 2} \label{Eq:su3_hoping_range_asymp}
\end{eqnarray}

In the later section, we will show the existence of the finite
excitation gap within the single-mode approximation.  In the
presence of the energy gap, the uniqueness of the ground state
implies that it is an incompressible quantum liquid.  Here we
shall consider the more general case.

Our current formalism can be generalized to the SU(N) case very
easily.  The SU(N) fundamental spinor given by
\begin{eqnarray}
\psi_\alpha(\{z_i\})=\left( \begin{array}{c} u_1 \\ u_2 \\ \cdot \\ \cdot \\ u_N \end{array}\right)=\frac{1}{\sqrt{1+\sum_{i=1}^N\bar{z}_iz_i}}\left(\begin{array}{c}1 \\
z_1
\\ z_2 \\ \cdot\\\cdot\\ z_N \end{array} \right), \label{Eq:suN_uvw}
\end{eqnarray}
can be used to parameterized CP$^{N-1}$, where $z_i$ are complex
numbers.  The SU(N) coherent states given by
\begin{eqnarray}
\sqrt{\frac{p!}{\prod_{i=1}^{N}m_i!}}u_1^{m_1}u_2^{m_2}\cdot\cdot
u_N ^{m_N}, \ \sum_{i=1}^N m_i\!=\!p \label{Eq:suN_coherent_state}
\end{eqnarray}
are described by the SU(N) $(p,0,0,..,0)$ multiplet.   A general
SU(N) multiplet is labelled by $N-1$ integers $(n_1,..,n_N)$. The
rank of the SU(N) group is $N-1$. Therefore,
Eq.(\ref{Eq:suN_coherent_state}) represents a $(N-1)$-dimensional
lattice.  As we have seen, it is a triangular lattice in two
dimensions, and in three dimensions it is a tetrahedral lattice, and so on.
One of the common features is that they are all frustrated
lattices.  The number of the lattice is also related to $p$,
namely the maximum power of the $u_i$, given by the following
formula
\begin{eqnarray}
d_N(p)=\frac{1}{(N-1)!}(p+1)(p+2)\cdot\cdot(p+N-1).
\label{Eq:suN_dim}
\end{eqnarray}
When the number of the electrons is equal to the number of sites,
$N=d_N(p)$, the fermionic many-body wavefunction is given by
\begin{eqnarray}
\Psi_N = \left| \begin{array}{ccccc} u_{11}^p & u_{11}^{p-1}u_{21}
&
. & . & u_{N1}^p \\ u_{12}^p & u_{12}^{p-1}u_{22} & . & . & u_{N2}^p \\
. & . & . & . & .
\\ . & . & . & . & . \\ u_{1N}^p & u_{1N}^{p-1}u_{2N} & . & . & u_{NN}^p
\end{array} \right| \label{Eq:suN_IQHE}
\end{eqnarray}
where $u_{ij}$ is the $i^{\text{th}}$ component of spinor in
Eq.(\ref{Eq:suN_uvw}) for the $j^{\text{th}}$ electron. Similarly,
let us consider the wavefunction $\Psi^m_N$.  Because the maximum
power of $u_1$ for each electron becomes $mp$, it describes a
state with fractional filling factor
\begin{eqnarray}
\nu_N=\frac{(p+1)(p+2)\cdot\cdot(p+N-1)}{(mp+1)(mp+2)\cdot\cdot(mp+N-1)}
\end{eqnarray}
which approaches to $1/m^{N-1}$ or $1/m^D$, where $D$ is the dimensionality, in the thermodynamic limit.

The Hamiltonian for the $\Psi^m_N$ is the zero-energy state is
given by
\begin{eqnarray}
H=\frac{1}{2}\sum_{i\ne j}\sum_{q=1, \ \text{odd}}^{q\leq m\!-\!2}
\kappa_q~P_{ij}^{(2mp\!-\!2q,q,0,0,..,0)_N}. \label{Eq:suN_H}
\end{eqnarray}
where $\kappa_q$ are positive and
$P_{ij}^{(2mp\!-\!2q,q,0,0,..,0)_N}$ projects two SU(N)
$(mp,0,..,0)_N$ spins to the $(2mp\!-\!2q,q,0,0,..,0)_N$ spin. The
argument for $\Psi^m_N$ to be the zero-energy state is quite
similar to the SU(3) case. We also leave the proof for the
non-degeneracy of the ground state in the Appendix.

Each the SU(N) coherent state is non-degenerate.  States in the
SU(N) $(p,0,..,0)_N$ multiplet can be labelled by its SU(N-1)
subgroup following the relation
\begin{eqnarray}
(p,0,0,..,0)_N=\bigoplus_{x=0}^{p} (x,0,..,0)_{N-1}
\label{Eq:suN_decomp}
\end{eqnarray}
From Eq.(\ref{Eq:su3_su_2decom}) and Eq.(\ref{Eq:suN_decomp}),
there is an inductive relation to label the states in the SU(N)
$(p,0,..,0)_N$ multiplet.  In other words, they can be labelled by
the SU(N-1), SU(N-2), .., and SU(2) subgroups namely
\begin{eqnarray}
|(mp,0,0,..,0)_N;x_1,x_2,..,x_{N\!-\!1}>
\end{eqnarray}
where $x_i$ denote the SU($i$) $(x_i,0,..,0)_i$ multiplets. Using
this coordinate system, the electron creation operator is defined
by
\begin{eqnarray}
\!\!\!\!|(p,0,0,..,0)_N;x_1,x_2,..,x_{N\!-\!1}\!\!>\to\! c^\dag_{(x_1,x_2,..,x_{N\!-\!1})}|0\!\!>
\end{eqnarray}
Let us denote $\vec{x}=(x_1,x_2,..,x_{N\!-\!1})$.  Then, similar
to the Eq.(\ref{Eq:su3_H}), Eq.(\ref{Eq:suN_H}) can be written as
\begin{eqnarray}
\!\!\!\!\!\!H\!\!=\!\kappa\!\!\!\sum_{\vec{R}\in(\!2\!m\!p\!-2,1,0,..,0\!)_{\!N\!}}\!\!\sum_{\vec{y},\vec{x}}F(\!\vec{R},\!\vec{y}\!)F(\!\vec{R},\!\vec{x}\!)c^{\dag}_{\vec{y}}c^{\dag}_{\vec{R}\!-\!\vec{y}}c_{\vec{R}\!-\!\vec{x}}c_{\vec{x}}
\label{Eq:suN_second_H}
\end{eqnarray}
for $m=3$, where $F(\vec{R},\vec{x})$ is the Clebsh-Gordan
coefficient from two SU(N) $(3p,0,0,..,0)_N$ spins to the
$(6p-2,1,0,..,0)_N$ spin.  The only tricky point in
Eq.(\ref{Eq:suN_second_H}) is the center-of-mass $\vec{R}$.
Because the SU(N) $(6p-2,1,0,0,..,0)_N$ multiplet contains
degeneracy in its states, the $(N-1)$-dimensional vector $\vec{R}$
is the function of the quantum numbers of the subgroups and the
Cartan subalgebra. Hence, we have shown that the general
Hamiltonian so that the general Laughlin states are the
non-degenerate ground state.  It is obvious to see that it
preserves the conservation law of the center-of-mass position.

So far there is no efficient way to calculate the Clebsh-Gordan
coefficients for the general SU(N)$\times$SU(N) to SU(N) case.  In
the SU(3) case, only the Clebsh-Gordan coefficients for small
multiplets can be calculated numerically.  However,
$F(\vec{R},\vec{x})$ can be obtained analytically in general
because
\begin{eqnarray}
(mp,0,0,..,0)_N\times(mp,0,0,..,0)_N\nonumber
\\=\bigoplus_{q=1}^{mp} (2mp-2q,q,0,..,0)_N,
\label{Eq:suN_CG}
\end{eqnarray}
where the right hand side contains no repeated multiplets.  We
leave this important mathematical problem to mathematical
physicists and conjecture that the pair-hopping range defined by
the half width of the $|F(\vec{R},\vec{x})|^2$ is also long-ranged
scaling as $\sqrt{p}$ with an exponential decay tail for the
general case.

\section{Single mode approximation} \label{section:sma}

In the SU(2) case, Girvin et al.\cite{girvin1985, girvin1986}
showed that for any liquid ground state in the lowest Landau
level, the density fluctuation costs finite energy, which implies
incompressibility, within the single mode approximation.  In this
section, we generalize their result to the SU(N) model.

As mentioned in the earlier section, our lattice model corresponds
to the fractional quantum Hall effect in CP$^{N-1}$ which is
subject to the background U(1) magnetic field with the
quantization $n$. The single-particle Lagrangian in the lowest
Landau level is given by
\begin{eqnarray}
L=-in\bar{\psi}_{\alpha}\frac{d\psi_{\alpha}}{dt}
\label{Eq:larangian}
\end{eqnarray}
Using Eq.(\ref{Eq:suN_uvw}), in the flat-space limit
Eq.(\ref{Eq:larangian}) can be written as
\begin{eqnarray}
L=\sum_{k=1}^{N-1} nx_k\dot{y}_k-ny_k\dot{x}_k=A_j\dot{X}_j
\label{Eq:flat_lag}
\end{eqnarray}
where we set $|z_k|\ll 1$ and $z_k=x_k+iy_k$ for $k=1$ to $N-1$
and $A_j$ and $X_j$ are the $2(N-1)$-dimensional vector potential
and the position vector $(\{x_k,y_k\})$ respectively. From
Eq.(\ref{Eq:flat_lag}), the single-particle orbit in the lowest
Landau level can be obtained as
\begin{eqnarray}
\Phi_{\{l_k\}}(\{z_k\})=\prod_{k=1}^{N-1}\frac{1}{\sqrt{2\pi
2^{l_k}l_k!}}z^{l_k}_ke^{\frac{-|z_k|^2}{4}} \label{Eq:lll_N}
\end{eqnarray}
where $l_k$ are non-negative integers.  We recognize that
Eq.(\ref{Eq:lll_N}) is the product of the $(N-1)$
lowest-Landau-level wavefunctions in two space dimensions.  Thus,
in the flat-space limit, the lowest Landau level in CP$^{N-1}$
becomes the direct product of $(N-1)$ quantum Hall effects in two
dimensions.  It can be seen by the non-commutative algebra as
well. From Eq.(\ref{Eq:flat_lag})
\begin{eqnarray}
[x_k,y_k ]=-\frac{i}{n} \ \text{for} \ k=1 \ \text{to} \ N-1
\label{Eq:nc}
\end{eqnarray}
which indicates the there are $N-1$ independent non-commutative
planes.  In the rest of the section, we present the results for
$N=3$.  The formalism can be generalized to any $N$ easily.  We
also note that in the rest of the section $N$ is the symbol for
the number of particles.

In the single-mode approximation (SMA), the variational excitation
energy is given by
\begin{eqnarray}
\Delta(k_1,k_2)=f(k_1,k_2)/s(k_1,k_2),
\end{eqnarray}
where $k_i$ are the complex wave numbers in the $i^{\text{th}}$
quantum Hall plane and $f(k_1,k_2)$ and $s(k_1,k_2)$ are the
oscillator strength and the static autocorrelation function given
by
\begin{eqnarray}
&&f(k_1,k_2)=\frac{1}{N}<\Psi_m|[\rho^\dagger_{(k_1,k_2)},[H,\rho_{(k_1,k_2)}]]|\Psi_m> \\
&&s(k_1,k_2)=\frac{1}{N}<\Psi_m|\rho^\dagger_{(k_1,k_2)}\rho_{(k_1,k_2)}|\Psi_m>
\end{eqnarray}
respectively, where $\rho_{(k_1,k_2)}$ is the density operator.
Here we use adopt Girvin et al.'s
notation\cite{girvin1985,girvin1986} which is a little bit
different from our previous Letter\cite{Chern2007PRL}.  In the
lowest Landau level, both $f(k_1,k_2)$ and $s(k_1,k_2)$ should be
treated carefully because coordinates do not mutually commute.
Particularly, the kinetic energy vanishes and they should be
redefined by
\begin{eqnarray}
\bar{f}(k_1,k_2)=\frac{1}{N}<\Psi_m|[\bar{\rho}^\dagger_{(k_1,k_2)},[\bar{V},\bar{\rho}_{(k_1,k_2)}]]|\Psi_m\!\!> \\
\bar{s}(k_1,k_2)=\!\frac{1}{N}\!<\!\!\Psi_m|\bar{\rho}^\dagger_{(k_1,k_2)}\bar{\rho}_{(k_1,k_2)}|\Psi_m\!\!>~~~~~~~~
\end{eqnarray}
where $\bar{\rho}$ is the projected density operator and $\bar{V}$
is the projected potential energy in the lowest Landau level,
which are respectively given by
\begin{eqnarray}
\bar{\rho}_{(k_1,k_2)}=\sum_{j=1}^{N}e^{-ik_1\frac{\partial}{\partial
z_{1j}}}e^{-ik_2\frac{\partial}{\partial
z_{2j}}}e^{-\frac{ik^*_2}{2}z_{2j}}e^{-\frac{ik^*_1}{2}z_{1j}}&&
\nonumber \\
\!
\bar{V}\!\!=\!\!\frac{1}{2}\!\int\!\frac{d^2\!q_1\!d^2\!q_2\!}{(2\pi)^4}v(\!q_1\!,q_2\!)(\bar{\rho}^\dag_{(\!q_1,q_2\!)}\bar{\rho}_{(\!q_1,q_2\!)}\!-\!\rho
e^{-\!\frac{|\!q_1\!|^2\!+\!|\!q_2\!|\!^2}{2}} )&&
\label{projected_V}
\end{eqnarray}
where $\rho$ is the average density.  In Eq.(\ref{projected_V}),
$v(q_1,q_2)$ is required to be positive indicating the repulsive
interaction to ensure the excitation energy to be positive.  Using
the algebra for the density operator
\begin{eqnarray}
\!\!\!\!\![\bar{\rho}_{(k_1,k_2)},\bar{\rho}_{(q_1,q_2)}]\!=\!(e^{\frac{k_1^*\!q_1\!+\!k_2^*\!q_2}{2}}\!\!\!-\!\!e^{\frac{k_1\!q^*_1\!+\!k_2\!q^*_2}{2}}\!)\bar{\rho}_{(k_1\!+\!q_1\!,k_2\!+\!q_2\!)},~~
\end{eqnarray}
$\bar{f}(k_1,k_2)$ can be easily computed as
\begin{eqnarray}
\bar{f}(k_1,k_2)=\frac{1}{2}\sum_{q_1,q_2}v(q_1,q_2)(e^{\frac{q^*_1k_1+q^*_2k_2}{2}}-e^{\frac{q_1k^*_1+q_2k^*_2}{2}})
[\nonumber\\\bar{s}(q_1,q_2)e^{-\frac{|k_1|^2+|k_2|^2}{2}}(e^{-\frac{k^*_1q_1+k^*_2q_2}{2}}-e^{-\frac{k_1q^*_1+k_2q^*_2}{2}})\nonumber\\+\bar{s}(k_1\!+\!q_1\!,k_2\!+\!q_2)(e^{\frac{k^*_1q_1+k^*_2q_2}{2}}\!-\!e^{\frac{k_1q^*_1+k_2q^*_2}{2}})]~~~.
\end{eqnarray}
A direct expansion shows that $\bar{f}(k_1,k_2)$ vanishes in the
fourth order in $k$.  To show the necessary condition for
existence of the excitation gap, we have to demonstrate that
$\bar{s}(k_1,k_2)$ vanishes in the same order in $k$.  Then,
because of the isotropy between $k_1$ and $k_2$
\begin{eqnarray}
\Delta(k_1,k_2)=\frac{a|k_1|^4+b|k_1|^2|k_2|^2+a|k_2|^4}{c|k_1|^4+d|k_1|^2|k_2|^2+c|k_2|^4}
\label{Eq:del}
\end{eqnarray}
remains finite as $k$ approaches to zero in any direction, where
$a$, $b$, $c$ , and $d$ are constants.

The asymptotic behavior of $\bar{s}(k_1,k_2)$ can be analyzed by
relating with the radial distribution function $g(\vec{r})$ by
\begin{eqnarray}
s(\vec{k})=1+\rho\int
d^4re^{-i\vec{k}\cdot\vec{r}}[g(\vec{r})-1]+\rho(2\pi)^4\delta^4(\vec{k})~~
\end{eqnarray}
where $\vec{k}=(\text{Re}k_1,\text{Im}k_1 ,\text{Re}k_2
,\text{Im}k_2)$ and $\vec{r}=(x_1,y_1,x_2,y_2)$ are real vectors.
For filling factor $\nu=1/m^2$ we obtained
\begin{eqnarray}
\rho[g(\vec{r})-1]=\frac{m^2}{4\pi^2}\sum_{l_1,l_2=0}^{\infty}\frac{(\frac{r_1^2}{2})^{l_1}(\frac{r_2^2}{2})^{l_2}}{l_1!l_2!}e^{-\frac{r_1^2+r_2^2}{2}}(\nonumber\\
<n_{l_1l_2}n_{00}>-<n_{l_1l_2}><n_{00}>-\frac{1}{m^2}\delta_{(l_1l_2)(00)})~~
\end{eqnarray}
where $r_1^2=x_1^2+y_1^2$ and $r_2^2=x_2^2+y_2^2$, and
$n_{l_1l_2}=c^\dag_{l_1l_2}c_{l_1l_2}$ where $c^\dag_{l_1l_2}$ is
the electron creation operator in the orbit $(l_1,l_2)$. Establish
the relation between $\bar{s}(k_1,k_2)$ and $s(k_1,k_2)$
\begin{eqnarray}
\bar{s}(k_1,k_2)=s(k_1,k_2)-(1-e^{-\frac{k_1^2+k_2^2}{2}}),
\end{eqnarray}
and define
\begin{eqnarray}
M_{n_1n_2}\equiv \int d^2r_1d^2r_2
(\frac{r_1^2}{2})^{n_1}(\frac{r_2^2}{2})^{n_2}\rho[g(r_1,r_2)-1]~
\end{eqnarray}
One can compute easily that
\begin{eqnarray}
M_{00}=m^2(<Nn_{00}>-<N><n_{00}>)-1&& \nonumber
\\ M_{10}=m^2(<\!(L_1\!+\!N\!)\!n_{00}\!>\!-\!<\!L_1\!+\!N\!><\!n_{00}\!>\!)\!-\!1&&\nonumber \\
M_{01}=m^2(<\!(L_2\!+\!N\!)\!n_{00}\!>\!-\!<\!L_2\!+\!N\!><\!n_{00}\!>\!)\!-\!1&&~~~
\end{eqnarray}
where $N=\sum_{l_1l_2}n_{l_1l_2}$ is the total number of particles
and $L_i=\sum_{l_1l_2}(l_i)n_{l_1l_2}$ are the total angular
momentum on the $i^{\text{th}}$ quantum Hall plane. Because of the
conservation of the number of particles and the angular momentum,
their fluctuation is zero, so $M_{00}=M_{10}=M_{01}=-1$.  Then,
the second order of $k$ in $\bar{s}(k_1,k_2)$ vanishes.  The
asymptotic behavior of $\bar{s}(k_1,k_2)$ indeed scales as the
fourth order in $k$.  Eq.(\ref{Eq:del}) holds true.

The current analysis relies on the transformation from CP$^{N-1}$
to R$^{2(N-1)}$, namely the flat-space limit.  The former one is a
compact space whose volume is finite without boundary.  The later
one is a non-compact space whose volume is infinite but the
exponential term in Eq.(\ref{Eq:lll_N}) sets the natural boundary.
An importance question is whether the structure of the energy
spectrum is preserved in the transformation.  We believe that the
energy spectrum is not a one-to-one mapping, because in
R$^{2(N-1)}$ there are certainly gapless edge excitations.  On the
other hand, in CP$^{N-1}$ there is no edge excitation due to the
lack of the boundary.  However, the structure of the bulk
excitation is preserved because we do not introduce any flux which
generates the electro-motive force to close/open a gap in the
transformation. Therefore, the existence of the finite excitation
gap in R$^{2(N-1)}$ implies that our lattice model also has a
finite excitation gap.  Besides the edge modes, whether or not
there are other gapless bulk excitations can not be answered by
the current approximation in R$^{2(N-1)}$.  If there are gapless
bulk excitations in R$^{2(N-1)}$, it should be also true in our
lattice model.  As far as we can say, within the single mode
approximation, we conclude that our unique ground state describes
an incompressible quantum liquid.

\section{Discussion and Conclusion}\label{section:dis_con}
Guided by the new symmetry of the center-of-mass position, we construct a family of the models to describe the fractionally-filled incompressible liquid in \emph{any} dimension.  They are long range pair hopping model in the frustrated lattice in $d \ge 2$.  We prove rigorously the uniqueness of the ground state in the open boundary condition using the group-theoretical method.  We also compute the energy gap using the single mode approximation, which is still the best analytical method to show the finiteness of the energy gap in the fractional quantum Hall effect.  Since the model is highly related to the higher dimensional generalization of the quantum Hall effect, one can generalize our models in different topological structures, for example, the torus and discuss the possibility of the fractionally-charged excitation for the future explorations.

We dedicate this work to Darwin Chang who made important
contributions at the early stage of  this work, but passed away
before its completion.  A special and deep gratitude should give to Dung-Hai Lee for leading the author to this topic.  We are supported by the NSC 97-2112-M-002-027-MY3 of Taiwan.

\appendix
\section{Proof of the non-degeneracy of the zero-energy state}
In this section, we present the rigorous proof of the uniqueness
of the $\Psi_m$ as the zero-energy state of the Hamiltonian in the
Eq.(\ref{Eq:suN_H}). The key procedure has been outlined in our
previous Letter\cite{Chern2007PRL}.

For simplicity, we focus on $m=3$ and set $p$ to be an odd
integer.  In this case, Eq.(\ref{Eq:suN_H}) has the following form
\begin{eqnarray}
H=\frac{\kappa_1}{2}\sum_{i\neq j}P_{ij}^{(6p-2,1,0,0,..,0)}
\label{Eq:appendix_H_1}
\end{eqnarray}
The Hamiltonian contains the term which projects two SU(N) spins
in the $(3p,0,..,0)_N$ to the $(6p-2,1,0,..,0)_N$ subspace. The
direct product of two SU(N) $(3p,0,..,0)_N$ spins can be
decomposed by the SU(N) subspace given in the
Eq.(\ref{Eq:suN_CG}). Particularly, when considering the
antisymmetric combination, the complete set of spaces reduces to
\begin{eqnarray}
(mp,0,0,..,0)_N\times(mp,0,0,..,0)_N|_a\nonumber
\\=\bigoplus_{q=1,\ \text{odd}}^{mp} (2mp-2q,q,0,..,0)_N,
\label{Eq:suN_CG_A}
\end{eqnarray}
where only odd $q$ are allowed.  If $\chi$ is the zero-energy
state of Eq.(\ref{Eq:appendix_H_1}), there is no $(6p-2,1,0,..,0)$
component in its two-spin spectrum.  In other words,
\begin{eqnarray}
(P_{ij}^{(6p-6,3,0,..,0)_N}\!+\!P_{ij}^{(6p-10,5,0,..,0)_N}\!+..+\!P_{ij}^{(0,3p,0,..,0)})\chi=\chi\nonumber
\\ \label{Eq:appendix_H_2}
\end{eqnarray}
for any pair $(ij)$.  Eq.(\ref{Eq:appendix_H_2}) constraints the
symmetry properties in the zero-energy state.  We note that the
state in the SU(N) $(3p,0,..,0)_N$ multiplet can be written as the
symmetric product of $3p$ SU(N) fundamental spinor in the
Eq.(\ref{Eq:suN_uvw}).  For particle $j$, it is given by
\begin{eqnarray}
\psi_j^{\alpha_{j1}}\psi_j^{\alpha_{j2}}\psi_j^{\alpha_{j3}}..\psi_j^{\alpha_{j,3p}}
\label{Eq:appedix_single}
\end{eqnarray}
where ${\alpha_{jk}}$ runs from $1$ to $N$.
Eq.(\ref{Eq:appedix_single}) is the alternative way of expressing
Eq.(\ref{Eq:suN_coherent_state}).  Using
Eq.(\ref{Eq:appedix_single}) as the basis, $\chi$ in general can
be written as
\begin{eqnarray}
\chi=\sum_{\{\alpha_{jn}=1\}}^N\!\! C(\!\{\alpha_{jn}\!\})\!
\prod_{j=1}^{d(p)}\prod_{n=1}^{3p}\psi_j^{\alpha_{jn}}.
\label{Eq:appendix_psi}
\end{eqnarray}
where the wavefunction $C(\{\alpha_{jn}\})$ satisfies the
following Schr\"odinger equation from the
Eq.(\ref{Eq:appendix_H_2})
\begin{eqnarray}
&&C(..,\!\{\alpha_{i}\!\}..,\!\{\alpha_{j}\},..)\nonumber \\
&&=\!\!\!\sum_{\{\beta_{i}\}, \{\beta_{j}\}}\![
A_3(\{\alpha_{i}\},\!\{\alpha_{j}\};\!\{\beta_{i}\},\!\{\beta_{j}\})\nonumber
\\&&+
A_5(\{\alpha_{i}\},\!\{\alpha_{j}\};\!\{\beta_{i}\},\!\{\beta_{j}\})\!+...\nonumber\\&&
+A_{3p}(\{\alpha_{i}\},\{\alpha_{j}\},\{\beta_{i}\},\{\beta_{j}\})]
C (..,\{\beta_{i}\},..,\{\beta_{j}\},..)\nonumber \\
\label{Eq:appendix_suN}
\end{eqnarray}
for any pair $(ij)$, where $A_q$ are the tensors for the
projection operator $P_{ij}^{(6p-2q,q,0,..,0)_N}$. $A_q$ does the
following symmetric operations
\begin{enumerate}
\item $q$ of the $3p$ indices of particle $i$ is made antisymmetric to
$q$ indices of particle $j$.
\item the rest of the indices of particle $i$ is made totally
symmetric to the rest of the indices of particle $j$.
\end{enumerate}
There is degree of freedom to choose which pair of indices is made
antisymmetric in the symmetry operation given above.  For example,
$A_3$ can be written as
\begin{eqnarray}
&&A_3(\{\alpha_{i}\},\{\alpha_{j}\};\{\beta_{i}\},\{\beta_{j}\})\nonumber\\&&=\!\frac{1}{N_3}(\delta_{\beta_{i1}}^{\alpha_{i1}}\delta_{\beta_{j1}}^{\alpha_{j1}}
-\delta_{\beta_{j1}}^{\alpha_{i1}}\delta_{\beta_{i1}}^{\alpha_{j1}})(\delta_{\beta_{i,p+1}}^{\alpha_{i,p+1}}\delta_{\beta_{j,p+1}}^{\alpha_{j,p+1}}
-\delta_{\beta_{j,p+1}}^{\alpha_{i,p+1}}\delta_{\beta_{i,p+1}}^{\alpha_{j,p+1}})\nonumber\\&&(\delta_{\beta_{i,2p+1}}^{\alpha_{i,2p+1}}\delta_{\beta_{j,2p+1}}^{\alpha_{j,2p+1}}
-\delta_{\beta_{j,2p+1}}^{\alpha_{i,2p+1}}\delta_{\beta_{i,2p+1}}^{\alpha_{j,2p+1}})(\delta_{\beta_{i2}}^{\alpha_{i2}}
..\delta_{\beta_{ip}}^{\alpha_{ip}}\delta_{\beta_{i,p+2}}^{\alpha_{ip+2}}..\delta_{\beta_{i,2p}}^{\alpha_{i,2p}}\nonumber\\&&\delta_{\beta_{i,2p+2}}^{\alpha_{i,2p+2}}..\delta_{\beta_{i,3p}}^{\alpha_{i,3p}}\delta_{\beta_{j2}}^{\alpha_{j2}}
..\delta_{\beta_{jp}}^{\alpha_{jp}}\delta_{\beta_{j,p+2}}^{\alpha_{jp+2}}..\delta_{\beta_{j,2p}}^{\alpha_{j,2p}}\delta_{\beta_{j,2p+2}}^{\alpha_{j,2p+2}}..\delta_{\beta_{j,3p}}^{\alpha_{j,3p}}
\nonumber \\&&+\ \text{sym.}),
\end{eqnarray}
where $N_3$ is the normalization constant.  Consequently, for a
particular pair $(ij)$, one can arrange the antisymmetric pairs so
that $C$ becomes $-C$ by the following independent exchanges
\begin{eqnarray}
&&(\alpha_{i1}..\alpha_{ip})\leftrightarrow
(\alpha_{j1}..\alpha_{jp}) \nonumber \\
&&(\alpha_{i,p+1}..\alpha_{i,2p})\leftrightarrow
(\alpha_{j,p+1}..\alpha_{j,2p}) \nonumber \\
&&(\alpha_{i,2p+1}..\alpha_{i,3p})\leftrightarrow
(\alpha_{j,2p+1}..\alpha_{j,3p}) \label{Eq:appendix_sym}
\end{eqnarray}
If the symmetry property of $C$ shown above can be made true
\emph{simultaneously} for \emph{any} pair $(ij)$.  The argument
that $\Psi_m$ is the unique zero-energy state can be given as the
following.  Let us consider the independent exchange of the first
group of $p$ indices while keeping others fixed.  $C$ becomes $-C$
has to be established in any pair $(ij)$.  Because the number of
particle is $N=d(p)$ is exactly equal to the total number of
states that $p$ indices represents, $C$ is proportional to
\begin{eqnarray}
C \sim \epsilon\{(\alpha_{j1}..\alpha_{jp})\}
\end{eqnarray}
where $\epsilon\{(\alpha_{j1}..\alpha_{jp})\}$ is the rank $d(p)$
tensor with respect to the group exchange.  Similarly, the second
and third properties in the Eq.(\ref{Eq:appendix_sym}) leads to
\begin{eqnarray}
C \sim
\epsilon\{(\alpha_{j1}..\alpha_{jp})\}\epsilon\{(\alpha_{j,p+1}..\alpha_{j,2p})\}\epsilon\{(\alpha_{j,2p+1}..\alpha_{j,3p})\}\nonumber
\\ \label{Eq:appendix_C}
\end{eqnarray}
Using Eq.(\ref{Eq:appendix_C}), any zero-energy state $\chi$ is
proportional to $\Psi_m$.

Now, we shall prove that Eq.(\ref{Eq:appendix_sym}) can indeed be
made true for all pairs $(ij)$ simultaneously. Let us assume that
there exists a ground state solution whose $C$ does not satisfy
Eq.(\ref{Eq:appendix_sym}) for pair $(kl)$. It means there is at
least one group exchange, say
$\{\alpha_{k1}..\alpha_{kp}\}\leftrightarrow
\{\alpha_{l1}..\alpha_{lp}\}$, so that $C$ does not follow
Eq.(\ref{Eq:appendix_sym}). However, since the wavefunction still
has to satisfy Eq.(\ref{Eq:appendix_suN}) for $(k,l)$, one can
write
\begin{eqnarray}
C=\sum_{q=3, \text{odd}}^{3p} C_q,
\end{eqnarray}
where $C_q$ is the component of $C$ that is odd with respect to
exchange of {\it exactly} $q$ pair of indices between particle $k$
and $l$ and even with respect to the exchange of the rest. Now let
us consider the effect of
$\{\alpha_{k1}..\alpha_{kp}\}\leftrightarrow
\{\alpha_{l1}..\alpha_{lp}\}$ on $C$.  After the exchange, $C_q$
can either change sign or stay invariant depending on whether an
odd or even number (out of $q$) antisymmetric indices are
contained in the specified triplets. In other words upon
$\{\alpha_{k1}..\alpha_{kp}\}\leftrightarrow
\{\alpha_{l1}..\alpha_{lp}\}$ we have
\begin{eqnarray}
C\rightarrow \sum_{q=3,\text{odd}}^{3p}\eta_q C_q,
\end{eqnarray}
where $\eta_q=\pm 1$. Since Eq.(\ref{Eq:appendix_sym}) is not
satisfied, all $\eta_{q}$ must not simultaneously be $-1$. Now
consider a new $C$
\begin{eqnarray}
C'\equiv \frac{1}{2} \Big[C-\sum_{q=3,\text{odd}}^{3p}\eta_q
C_q\Big].
\end{eqnarray}
It is obvious that upon
$\{\alpha_{k1}..\alpha_{kp}\}\leftrightarrow
\{\alpha_{l1}..\alpha_{lp}\}$ $C'\rightarrow -C'$. Moreover by
construction $C'$ only contains those $C_q$ whose $\eta_q=-1$. Now
use $C'$ as the starting $C$ and repeat the above operation until
we reach a final $C'$ for which Eq.(\ref{Eq:appendix_sym}) holds
for all triplet exchanges and for all $(ij)$. Since at each stage
of obtaining $C'$ certain $C_q$ are projected out, there must be
missing components in the final $C$.  However we have already
proven that any $C$ that satisfy Eq.(\ref{Eq:appendix_sym}) for
all $(ij)$ pair must lead to the solution $\chi\sim\Psi_m$.
However, $\Psi_m$ contains all components for all pair $(ij)$.
Consequently we have reached a contradiction. Therefore it must be
possible to make Eq.(\ref{Eq:appendix_sym}) hold true for all
pairs $(ij)$ for any ground state solution satisfying
Eq.(\ref{Eq:appendix_H_2}).

The proof can be generalized to any $m$ by assigning $m$ groups of
indices.  Thus, we have proven that $\Psi_m$ is the unique
zero-energy state of Eq.(\ref{Eq:suN_H}).

\section{Summary of SU(3) algebra and representation
theory}\label{Appendix:SU3}
\subsection{Algebra}
The SU(3) group is the one with which people are very familiar
besides SU(2).  This note will not be a thorough review  of Lie
algebra but focuses on what we shall need in the paper.  The
generators of SU(3)

\begin{eqnarray}
\nonumber & & T_1 \!=\! \frac{1}{2}\left(
\begin{array}{cccc}
               0  & 1 & 0   \\
               1  & 0 & 0  \\
               0  & 0 & 0  \end{array} \right), \
 T_2\!=\! \frac{1}{2}\left( \begin{array}{cccc}
               0  & -i & 0 \\
               i  & 0 & 0 \\
               0  & 0 & 0
              \end{array} \right), \\ \nonumber & &
 T_3 \!=\! \frac{1}{2}\left( \begin{array}{cccc}
                1  & 0 & 0 \\
               0  & -1 & 0 \\
               0  & 0 & 0
               \end{array} \right), \
 T_4 \!=\! \frac{1}{2}\left( \begin{array}{cccc}
                0  & 0 & 1 \\
               0  & 0 & 0 \\
               1  & 0 & 0
               \end{array} \right), \  \\
 \nonumber & & T_5 \!=\! \frac{1}{2}\left(
\begin{array}{cccc}
               0  & 0 & -i \\
               0  & 0 & 0 \\
               i  & 0 & 0
               \end{array} \right), \
 T_6\!=\! \frac{1}{2}\left( \begin{array}{cccc}
               0  & 0 & 0 \\
               0  & 0 & 1 \\
               0  & 1 & 0
               \end{array} \right), \\ \nonumber &&
 T_7 \!=\! \frac{1}{2}\left( \begin{array}{cccc}
               0  & 0 & 0 \\
               0  & 0 & -i \\
               0  & i & 0
               \end{array} \right), \
 T_8 \!=\! \frac{1}{\sqrt{12}}\left( \begin{array}{cccc}
               1  & 0 & 0 \\
               0  & 1 & 0 \\
               0  & 0 & -2
               \end{array} \right)
\end{eqnarray}
satisfy
\begin{eqnarray}
Tr(T_aT_b)=\frac{1}{2}\delta_{ab}
\end{eqnarray}
in the standard convention.  The Cartan subalgebra contains $T_3$
and $T_8$.  Denote them by $H_1$ and $H_2$ respectively.  The
simply roots of SU(3) can be obtained as
\begin{eqnarray}
\alpha^1=(\frac{1}{2}, \frac{\sqrt{3}}{2}),\
\alpha^2=(\frac{1}{2}, -\frac{\sqrt{3}}{2})
\end{eqnarray}
All positive roots of SU(3) are given by $\alpha^1$, $\alpha^2$,
and $\alpha^1+\alpha^2$.  The generators correspond to the
positive roots are given by
\begin{eqnarray}
E_{\alpha^1}=\frac{1}{\sqrt{2}}(T_4+iT_5),\nonumber\\
E_{\alpha^2}=\frac{1}{\sqrt{2}}(T_6-iT_7), \nonumber\\
E_{\alpha^1+\alpha^2}=\frac{1}{\sqrt{2}}(T_1+iT_2)
\end{eqnarray}
These generators are the raising operators in SU(3).  Their lowing
operators are the Hermitian conjugates of themselves, \emph{i.e.}
\begin{eqnarray}
E_{-\alpha^1}=\frac{1}{\sqrt{2}}(T_4-iT_5), \nonumber \\
E_{-\alpha^2}=\frac{1}{\sqrt{2}}(T_6+iT_7), \nonumber \\
E_{-\alpha^1-\alpha^2}=\frac{1}{\sqrt{2}}(T_1-iT_2)
\end{eqnarray}
The algebra is given by
\begin{eqnarray}
&& [E_{\alpha^1}, E_{-\alpha^1}]=E^1_3=\alpha^1\cdot H \label{Eq:su3_algebra1}\\
&& [E_{\alpha^2}, E_{-\alpha^2}]=E^2_3=\alpha^2\cdot H \label{Eq:su3_algebra2}\\
&& [E_{\alpha^1}, E_{\alpha^2}]=\frac{1}{\sqrt{2}}E_{\alpha^1+\alpha^2} \label{Eq:su3_algebra3}\\
&& [E_{-\alpha^1},
E_{\alpha^1+\alpha^2}]=\frac{1}{\sqrt{2}}E_{\alpha^2} \label{Eq:su3_algebra4}\\
&& [E_{-\alpha^2},
E_{\alpha^1+\alpha^2}]=-\frac{1}{\sqrt{2}}E_{\alpha^2}
\label{Eq:su3_algebra5}
\end{eqnarray}
The fundamental weight is defined by
\begin{eqnarray}
\frac{2\alpha^i\cdot \mu^j}{|\alpha^i|^2}=\delta_{ij}
\end{eqnarray}
In SU(3), $\mu^i$'s are given by
\begin{eqnarray}
\mu^1=(\frac{1}{2}, \frac{\sqrt{3}}{6}),\ \mu^2=(\frac{1}{2},
-\frac{\sqrt{3}}{6})
\end{eqnarray}
The representation whose highest weight is the fundamental weight
is called the fundamental representation.  Since the rank of SU(N)
group is $N-1$.  The number of simply root and that of the
fundamental weight are also $N-1$.  The highest weight $\mu$ in
\emph{any} SU(3) representation is given by $\mu=p\mu^1+q\mu^2$.
$p$ and $q$ are called Dynkin coefficients, which are unique for
every representation.  Therefore, SU(3) representations are
denoted by $(p,q)$.  The total number of the Casimir operators of
SU(N) is also equal to its rank.  We shall pay our attention to
the quadratic Casimir operator only.  It is defined by
\begin{eqnarray}
C=\sum_a T_aT_a
\end{eqnarray}
We can compute it to be
\begin{eqnarray}
C=H^2_1+H^2_2+\sum_{all\ positive\ roots}E_{\alpha}E_{-\alpha}
\end{eqnarray}
Using
Eq.(\ref{Eq:su3_algebra1})(\ref{Eq:su3_algebra2})(\ref{Eq:su3_algebra3})(\ref{Eq:su3_algebra4})(\ref{Eq:su3_algebra5})
and the highest weight method, we compute the quadratic Casimir
for SU(3) $(p,q)$ representation:
\begin{eqnarray}
C[p,q]=\frac{1}{3}(p^2+pq+q^2+3p+3q) \label{Eq:su3_casimir}
\end{eqnarray}
The dimension of $(p,q)$ representation is given by
\begin{eqnarray}
D[p,q]=\frac{(p+1)(q+1)(p+q+2)}{2} \label{Eq:su3_dimension_2}
\end{eqnarray}
In the context, we introduce the SU(3) algebra in favor of
particle physics.  Namely, the SU(3) algebra is given in the
Gell-Mann notation.  There is another basis which is also very
interesting and useful in certain problems \cite{swart1963RMP}.

If we define
\begin{eqnarray}
(A^i_k)_{\mu\nu}=\delta_{i\nu}\delta_{k\mu}-\frac{1}{3}\delta_{ik}\delta_{\mu\nu}
\label{Eq:su3_okubo_notation_1}
\end{eqnarray}
where $i,k,\mu,\nu=1,2,3$, having the following properties:
\begin{eqnarray}
A^i_k=(A^k_i)^\dag \label{Eq:su3_okubo_notation_2}\\
A^1_1+A^2_2+A^3_3=0 \label{Eq:su3_okubo_notation_3}
\end{eqnarray}
in which we know that there are only 8 independent generators.  It
can be checked that they satisfy the following commutation
relations:
\begin{eqnarray}
[A^i_k, A^j_l]=\delta^i_l A^j_k - \delta^j_k A^i_l
\label{Eq:su3_okubo_notation_4}
\end{eqnarray}
These 8 independent generators form the SU(3) algebra.  The Cartan
subalgebra is given by
\begin{eqnarray}
h_1 &=& \frac{1}{2}(A^1_1-A^2_2) = T_3  \nonumber \\
h_2 &=& \frac{1}{2}(A^2_2 - A^3_3) =
-\frac{1}{2}T_3+\frac{\sqrt{3}}{2}T_8
\label{Eq:su3_okubo_notation_5}
\end{eqnarray}
This notation is so-called Okubo's notation.  The relation between
Okubo notation and the Gell-Mann notation is given as the
following.  Denote
\begin{eqnarray}
T_1&=&I_1, \ T_2=I_2,\  T_3=I_3 \nonumber\\
T_4 &=& K_1,\ T_5=K_2 \nonumber\\
T_6&=&L_1,\ T_7=L_2 \nonumber\\
T_8&=&M \label{Eq:su3_Gell-Mann_Okubo_1}
\end{eqnarray}
and
\begin{eqnarray}
I_\pm&=&I_1\pm i I_2 \nonumber \\
K_\pm &=& K_1 \pm iK_2 \nonumber \\
L_\pm &=& L_1 \pm iL_2, \label{Eq:su3_Gell-Mann_Okubo_2}
\end{eqnarray}
and the $A^i_k$ can be written as
\begin{eqnarray}
A^1_1&=&I_3+\frac{1}{3}\sqrt{3}M,\ A^2_1 = I_+,\ A^1_2=I_-,
\nonumber
\\ A^2_2&=&-I_3+\frac{1}{3}\sqrt{3}M,\ A^3_1=K_+,\ A^1_3=K_-,
\nonumber \\ A^3_3&=&-\frac{2}{3}\sqrt{3}M,\ A^3_2=L_+,\ A^2_3=L_-
\label{Eq:su3_Gell-Mann_Okubo_3}
\end{eqnarray}
On the other hand, it is possible to obtain the transformation
between $I$, $K$, and $L$: Define
\begin{eqnarray}
P_i=e^{i\pi I_2} \nonumber \\
P_k=e^{i\pi K_2} \nonumber \\
P_l=e^{i\pi L_2} \label{Eq:su3_Gell-Mann_transformation_1},
\end{eqnarray}
Then
\begin{eqnarray}
P^{-1}_iI_\pm P_i=-I_{\mp},\ P^{-1}_i K_\pm P_i = L_\pm, \nonumber
\\ P^{-1}_k I_\pm P_k = L_\mp, \ P^{-1}_k K_\pm P_k = -K_\mp,
\nonumber \\P^{-1}_l I_\pm P_l = K_\pm, P^{-1}K_\pm P_l = -I_\pm,
\nonumber \\ P^{-1}L_\pm P_i = - K_\pm, \ P^{-1}_k L_\pm P_k =
-I_\mp, \nonumber \\ P^{-1}_l L_\pm P_l = -L_\mp
\label{Eq:su3_Gell-Mann_transformation_2}
\end{eqnarray}
\subsection{Representation in the X-$\text{L}_3$ basis}
The X operator satisfies the following equation
\begin{eqnarray}
X|(p,q)jLL_z\rangle = 2p+q-6j |(p,q)jLL_z\rangle \label{X}
\end{eqnarray}
$j$ ranges from $0,\frac{1}{2}$,..to $\frac{p+q}{2}$.  $j=0$ is
the highest X state.  $j$ is a quantum number on the X-axis.
However, due to degeneracy, for a certain $j$, it could be many
$L$.  The value of $L$ is given by
$|j-\frac{q}{2}|$...$j+\frac{q}{2}$.  Using this basis the SU(3)
Clebsch-Gordan coefficients (isofactor) are given by
\begin{eqnarray}
&&\langle(p_1,0)M,(p_2,0)N||(p,q)jL\rangle \nonumber \\&=&
(-1)^q\sqrt{\frac{(p+1)(q+1)!(p+q+L-M-N+1)!}{(p_1-q)!(p_2-q)!}}
\nonumber\\&\times&\sqrt{\frac{(p+q-L-M-N)!(2M+2N-q+1)!}{(p_1-2M)!(p_2-2N)!}}
\nonumber\\&\times&
\frac{\sqrt{(2M+1)!(2N+1)!}}{(p+q+1)!}\nonumber \\
&\times&\sum_{I'_1+I'_2=\frac{q}{2}}(-1)^{2I'_2}\frac{(p_1-2I'_1)!(p_2-2I'_2)!}{(2I'_1)!(2I'_2)!(2M-2I'_1)!(2N-2I'_2)!}
\nonumber \\
&\times& \left\{\begin{array}{ccc} I'_1 & I'_2 & \frac{q}{2} \\
M-I'_1 & N-I'_2 & j \\ M & N & L
\end{array}\right\}
\end{eqnarray}
where $j=M+N-\frac{q}{2}$ and we use the Wigner 9-j symbol.  If
$L=M+N$, namely $L=j+\frac{q}{2}$ case, the Wigner 9-j symbol has
a simpler form:
\begin{eqnarray}
&&\left\{\begin{array}{ccc}I'_1 & I'_2 & \frac{q}{2}\\ M-I'_1 &
N-I'_2 & j \\ M & N & M+N
\end{array}\right\} \nonumber \\&=& \frac{1}{\sqrt{(q+1)(2M+1)(2N+1)(2j+1)}}
\end{eqnarray}
The CG coefficient becomes
\begin{eqnarray}
&&\langle(p_1,0)M,(p_2,0)N||(p,q)jL\rangle \nonumber \\&=&
(-1)^q\sqrt{\frac{(p+1)q!(p+q-2M-2N)!}{(p_1-q)!(p_2-q)!}}
\nonumber \\&\times&
\sqrt{\frac{(2M)!(2N)!(2M+2N-q)!}{(p+q+1)!(p_1-2M)!(p_2-2N)!}}
\nonumber\\&\times&
\sum_{I'_1+I'_2=\frac{q}{2}}(-1)^{2I'_2}\frac{(p_1-2I'_1)!(p_2-2I'_2)!}{(2I'_1)!(2I'_2)!(2M-2I'_1)!(2N-2I'_2)!}
\nonumber \\
\end{eqnarray}
This result can be compared with the CG coefficient to the highest
weight:
\begin{eqnarray}
&&\langle (p_1,0)M,(p_2)N || (p,q)0,\frac{q}{2}\rangle
\nonumber\\&=&
(-1)^{q+2M}\frac{(p+1)!q!}{(p+q+1)!(p_1-q)!(p_2-q)!} \nonumber
\\&\times& \frac{(p_1-2M)!(p_2-2N)!}{(2M)!(2N)!}\delta_{M+N,\frac{q}{2}}
\end{eqnarray}

%

\end{document}